\documentclass[12pt]{iopart}
\usepackage{iopams}
\usepackage[dvips]{graphicx}

\begin{document}

\title{Self-similarity, small-world,
scale-free scaling, disassortativity, and robustness in hierarchical
lattices}

\author{Zhongzhi Zhang}
\address{Department of Computer Science and Engineering, Fudan
University, Shanghai 200433, China}
\address{Shanghai Key Lab of Intelligent Information Processing,
Fudan University, Shanghai 200433, China} \ead{zhangzz@fudan.edu.cn}

\author{Shuigeng Zhou}
\address{Department of Computer Science and Engineering, Fudan
University, Shanghai 200433, China}
\address{Shanghai Key Lab of Intelligent Information Processing,
Fudan University, Shanghai 200433, China} \ead{sgzhou@fudan.edu.cn}

\author{Tao Zou}
\address{Department of Computer Science and Engineering, Fudan
University, Shanghai 200433, China}
\address{Shanghai Key Lab of Intelligent Information Processing,
Fudan University, Shanghai 200433, China}

\date{\today}
\begin{abstract}
In this paper, firstly, we study analytically the topological
features of a family of hierarchical lattices (HLs) from the view
point of complex networks. We derive some basic properties of HLs
controlled by a parameter $q$: scale-free degree distribution with
exponent $\gamma=2+{\frac{\ln 2}{\ln q}}$, null clustering
coefficient, power-law behavior of grid coefficient, exponential
growth of average path length (non-small-world), fractal scaling
with dimension $d_B=\frac{\ln (2q)}{\ln 2}$, and disassortativity.
Our results show that scale-free networks are not always
small-world, and support the conjecture that self-similar scale-free
networks are not assortative. Secondly, we define a deterministic
family of graphs called small-world hierarchical lattices (SWHLs).
Our construction preserves the structure of hierarchical lattices,
including its degree distribution, fractal architecture, clustering
coefficient, while the small-world phenomenon arises. Finally, the
dynamical processes of intentional attacks and collective
synchronization are studied and the comparisons between HLs and
Barab{\'asi}-Albert (BA) networks as well as SWHLs are shown. We
find that the self-similar property of HLs and SWHLs significantly
increases the robustness of such networks against targeted damage on
hubs, as compared to the very vulnerable non fractal BA networks,
and that HLs have poorer synchronizability than their counterparts
SWHLs and BA networks. We show that degree distribution of
scale-free networks does not suffice to characterize their
synchronizability, and that networks with smaller average path
length are not always easier to synchronize.

\end{abstract}

\maketitle


\section{Introduction}
Topological characteristics, such as scale-free degree distribution,
small-world effect, fractal scaling and degree correlations, have
recently attracted much attention in network science. The last few
years have witnessed a tremendous activity devoted to the
characterization and understanding of networked
systems~\cite{AlBa02, DoMe02, Ne03, BoLaMoChHw06}. Small-world
property~\cite{WaSt98} and scale-free behavior~\cite{BaAl99} are two
unifying concepts constituting our basic understanding of the
organization of real-life complex systems. Small-world property
refers to the one that the expected number of edges (links) needed
to pass from one arbitrarily selected node (vertex) to another one
is low, which grows at most logarithmically with the number of
nodes. Scale-free behavior means the majority of nodes in a network
have only a few connections to other nodes, whereas some nodes are
connected to many other nodes in the network. This poses a
fundamental question how these two characteristics are related. It
has been observed that small-world property and scale-free behavior
are not independent~\cite{CoHa03}: scale-free networks, normally,
have extremely short average path length (APL), scaling
logarithmically or slower with system size. Is this universal?

In fact, the above mentioned two properties (i.e. small-world
property and scale-free behavior) do not provide sufficient
characterizations of the real-world systems. It has been observed
that real networks exhibit ubiquitous degree correlations among
their
nodes~\cite{MsSn02,PaVaVe01,VapaVe02,Newman02,Newman03c,BoPa03,BoPa05}.
This translates in the observation that the degrees of nearest
neighbor nodes are not statistically independent but mutually
correlated in practically every network imaginable. Correlations
play an important role in the characterization of network topology,
and have led to a first classification of complex
networks~\cite{Newman02}. A series of recent measurements indicate
social networks are all assortative, while all biological and
technological networks disassortative: in social networks there is a
tendency for the hubs to be linked together, in biological and
technological network the hubs show the opposite tendency, being
primarily connected to less connected nodes. Correlations are now a
very relevant issue, especially in view of the important
consequences that they can have on dynamical processes taking place
on networks~\cite{BoPa02,MoVa03,VaMo03,EcGaMoVa05}.

Recently, it has been discovered by the application of a
renormalization procedure that diverse real networks, such as the
WWW, protein-protein interaction networks and metabolic networks,
exhibit fractal scaling and topological
self-similarity~\cite{SoHaMa05,SoHaMa06,St05,YoRaOr06,GoSaKaKi06}.
Fractal scaling implies that in a network the minimum number of
node-covering boxes $N_{B}$ of linear size $\ell_{B}$ scales with
respect to $\ell_{B}$ as a power-law $N_{B}$, with an exponent that
is given by a finite fractal dimension $d_{B}$~\cite{SoHaMa05}.
Self-similarity refers to the invariant scale-free distribution
probability to find a node with degree $k$, $P(k)\sim k^{-\gamma}$,
i.e. the exponent $\gamma$ remains the same under the
renormalization with different box sizes~\cite{SoHaMa05,ZhYaWa05a}.
In complex networks, fractality and self-similarity do not always
imply each other: a fractal network model is self-similar, while a
self-similar network is not always fractal~\cite{GoSaKaKi06}. One
can obtain the fractal dimension $d_{B}$ by measuring the ratio of
$N_{B}$ over the total number of nodes $N$ in the network, which
satisfies $N_{B}/N\thicksim \ell_{B}^{-d_{B}}$. After renormalizing
the networks, the degree $k_{B}(\ell_{B})$ of a node in the
renormalized network versus the largest degree $k_{hub}$ inside the
box that was contracted to one node with degree $k_{hub}$ in the
renormalization process exhibits a scaling behavior:
$k_{B}(\ell_{B})=s(\ell_{B}) k_{hub}$, where $ s(\ell_{B}) $ is
assumed to scales like $ s(\ell_{B}) \thicksim \ell_{B}^{-d_{k}}$
with $d_{k}$ being the degree exponent of the boxes. In self-similar
scale-free networks, the relation between the three indexes
$\gamma$, $d_{B}$ and $d_{k}$ satisfies $\gamma =1+ d_{B}/
d_{k}$~\cite{SoHaMa05}.¡¡

Correlations and topological fractality are important properties for
many real-world complex systems. Then a natural fundamental question
is raised how the two characteristics relate to each other. Recent
researches~\cite{SoHaMa06,YoRaOr06} have shown that self-similar
scale-free networks are not assortative, and the qualitative feature
of disassortativity is scale-invariant under renormalization.
Moreover, self-similarity and disassortativity of scale-free
networks make such networks more robust against a sinister attack on
nodes with large degree, as compared to the very vulnerable non
fractal scale-free networks~\cite{SoHaMa06}.

However, do small-world property and scale-free behavior always go
along? How do systems have evolved into self-similar disassortative
scale-free networks? How the dynamical processes such as intentional
attack and synchronization are influenced by the topological
fractality and disassortativity of scale-free networks? Such a
series of important questions still remain open. To relate these
questions, in this paper we therefore launch a study seeking a
better understanding of the relations among these topological
properties.

It is of interest to study above important questions with
deterministic methods. Because of their strong advantages,
deterministic network models have received much
attention~\cite{BaRaVi01, DoGoMe02, CoFeRa04, ZhRoZh06c, JuKiKa02,
RaSoMoOlBa02, RaBa03, AnHeAnSi05, DoMa05,ZhCoFeRo05, ZhRo05,
CoOzPe00, CoSa02, ZhRoGo05,ZhWaHuCh04}. First, the method of
generating deterministic networks makes it easier to gain a visual
understanding of how networks are shaped, and how do different nodes
relate to each other~\cite{BaRaVi01}; moreover, deterministic
networks allow to compute analytically their topological properties,
which have played a significant role, both in terms of explicit
results and a guide to and a test of simulated and approximate
methods~\cite{BaRaVi01, DoGoMe02, CoFeRa04, ZhRoZh06c, JuKiKa02,
RaSoMoOlBa02, RaBa03, AnHeAnSi05, DoMa05,ZhCoFeRo05, ZhRo05,
CoOzPe00, CoSa02, ZhRoGo05, ZhWaHuCh04}. On the other hand,
deterministic networks can be easily extended to produce random
variants which exhibit the classical characteristics of many
real-life
systems~\cite{DoMeSa01,OzHuOt04,ZhRoCo05a,ZhYaWa05,ZhRoCo05} .

Inspired by the above mentioned questions, here we first introduce a
deterministic family of networks. These networks are called
hierarchical lattices (HLs), which yield exact renormalization-group
solutions~\cite{BeOs79, KaGr81, GrKa82, Ya88, QiYa01, HiBe06}. From
the perspective of complex network, we show that HLs are
simultaneously scale-free, self-similar and disassortative, but lack
the small-world property. Then we present a deterministic
construction of a class of small-world hierarchical lattices
(SWHLs), which preserve the basic structure properties including
power law degree distribution, self-similarity, and
disassortativeness, while lead to the small-world effect. Finally,
we investigate the effects of network structures on the dynamics
taking place in them.

This article is organized as follows. In Sec.~II, we introduce the
construction of hierarchical lattices (HLs) and study their
topological features including the degree distribution, moments,
clustering coefficient, grid coefficient, self-similarity, degree
correlations, and average path length (APL).  The detailed exact
derivation about APL is shown in Appendix. In Sec.~III, we propose
the deterministic construction of the small-world hierarchical
lattices (SWHLs) and study their properties. In Sec.~IV, attack
tolerance of HLs is studied and the comparisons between HLs and
Barab{\'asi}-Albert (BA) networks are shown. In Sec.~V, we do a
comparative investigation of synchronization in HLs, SWHLs and BA
networks. Sec.~VI is devoted to our conclusions.

\section{Hierarchical lattices}
In this section, from topological perspective of complex networks,
we present the construction and the basic properties such as degree
distribution, clustering coefficient, average path length (APL),
fractality, and correlations of the hierarchical lattices  (HLs).

\begin{figure}
\begin{center}
\includegraphics[width=8cm]{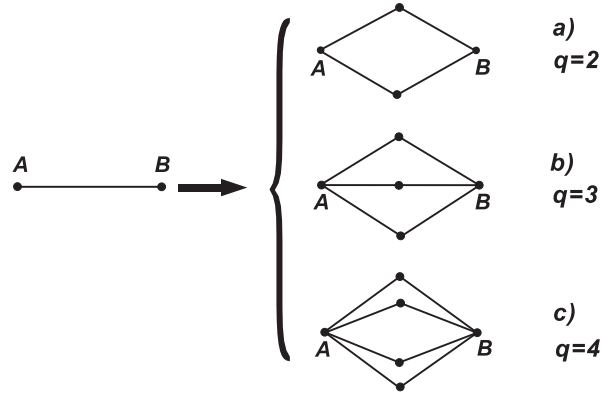}
\caption{Iterative construction method of the hierarchical lattices
for some limiting cases.} \label{fig1}
\end{center}
\end{figure}

\subsection{Construction of the Lattice}
\begin{figure}
\begin{center}
\includegraphics[width=8cm]{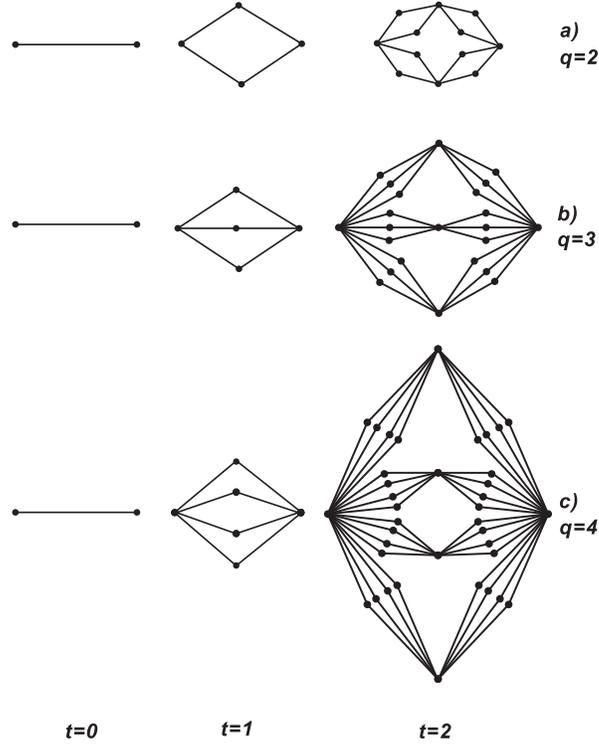}
\caption{Examples of the hierarchical lattices for some particular
case of $q=2$, $q=3$, and $q=4$, showing the first three steps of
the iterative process. } \label{fig2}
\end{center}
\end{figure}

The hierarchical lattices~\cite{Ya88} are constructed in an
iterative manner as shown in Fig.~\ref{fig1}. We denote the
hierarchical lattices (networks) after $t$ generations by $H(q,t)$,
$q\geq 2$ and $t\geq 0$.  The networks are constructed as follows:
for $t=0$, $H(q,0)$ is an edge connecting two points. For $t\geq 1$,
$H(q,t)$ is obtained from $H(q,t-1)$. We replace each of the
existing edges in $H(q,t-1)$ by the connected cluster of edges on
the right of Fig.~\ref{fig1}. The growing process is repeated $t$
times, with the infinite lattices obtained in the limit $t \to
\infty$. Figure~\ref{fig2} shows the growing process of the networks
for three particular cases of $q=2$, $q=3$, and $q=4$. It should be
noted that in the hierarchical lattice of $q=2$ case~\cite{BeOs79},
the Migdal-Kadanoff~\cite{MiEk7576,Ka76} recursion relations with
dimension $2$ and length rescaling factor $2$ are exact.

Griffiths and Kaufman provided two explanations for the construction
of the hierarchical lattices~\cite{GrKa82}, which are called
``aggregation'' and ``miniaturization''. In essence, these two
interpretations reflect the self-similar structure of the
hierarchical lattices, which allow one to calculate analytically
their topological characteristics.

Next we compute the numbers of nodes (vertices) and links (edges) in
$H(q,t)$. Let $L_v(t)$ and $L_e(t)$ be the numbers of vertices and
edges created at step $t$, respectively. Note that each of the
existing edges yields $q$ nodes, and the addition of each new node
leads to two new edges. By construction, for $t\geq 1$, we have
\begin{equation}
L_v(t)=qL_e(t-1)
\end{equation}
and
\begin{equation}
L_e(t)=2L_v(t).
\end{equation}
Considering the initial condition $L_v(0)=2$ and $L_e(0)=1$, it
follows that
\begin{equation}\label{Nv1}
L_v(t)=q\,(2q)^{t-1}
\end{equation} and
\begin{equation}\label{Ne1}
L_e(t)=(2q)^{t}.
\end{equation}
Thus the number of total nodes $N_t$ and edges $E_t$ present at step
$t$ is
\begin{eqnarray}\label{Nt1}
N_t=\sum_{t_i=0}^{t}L_v(t_i)=\frac{q(2q)^{t}+3q-2}{2q-1}
\end{eqnarray}
and
\begin{eqnarray}\label{Et1}
E_t=L_e(t)=(2q)^{t},
\end{eqnarray}
respectively.

\subsection{Degree distribution}
Let $k_i(t)$ be the degree of node $i$ at step $t$. Then by
construction, it is not difficult to find following relation:
\begin{equation}
k_i(t)=q\,k_i(t-1),
\end{equation}
which expresses a preference attachment \cite{BaAl99}. If node $i$
is added to the network at step $t_i$, $k_i(t_i)=2$ and hence
\begin{equation}
k_i(t)=2\,q^{t-t_i}\label{Ki1}.
\end{equation}
Therefore, the degree spectrum of the network is discrete. It
follows that the degree distribution is given by
\begin{equation}
P(k)=\left\{\begin{array}{lc} {\displaystyle{L_v(0)\over
N_t}={2\over \frac{q(2q)^{t}+3q-2}{2q-1}} }
& \ \hbox{for}\ t_i=0\\
{\displaystyle{L_v(t_i)\over N_t}={q\,(2q)^{t_i-1}\over
 \frac{q(2q)^{t}+3q-2}{2q-1}} }
& \ \hbox{for}\  t_i\ge 1\\
0 & \ \hbox{otherwise}\end{array} \right.
\end{equation}
and that the cumulative degree distribution \cite{Ne03,DoGoMe02} is
\begin{equation}
P_{\rm cum}(k)=\sum_{\rho \leq t_i}\frac{L_v(\rho)}{N_t}
={\frac{q(2q)^{t_i}+3q-2}{q(2q)^{t}+3q-2}}.
\end{equation}
Substituting for $t_i$ in this expression using $t_i=t-\frac{\ln
\frac{k}{2}}{\ln q}$ gives
\begin{eqnarray}\label{gamma1}
P_{\rm cum}(k)&=&{q\,(2q)^{t}\left({\frac{k}{2}}\right)^{-\frac{\ln(2q)}{\ln q}}+3q-2\over {q(2q)^{t}+3q-2}}\nonumber\\
          &\approx&\left({k\over2}\right)^{\textstyle-(1+{\ln 2\over \ln q})}\qquad\hbox{for large $t$}.
\end{eqnarray}
So the degree distribution follows a behavior power law with the
exponent $\gamma=2+{\frac{\ln 2}{\ln q}}$. For $q=2$, Eq.
(\ref{gamma1}) recovers the result of the particular case $p=0$
previously obtained in Ref. \cite{DoGoMe02}.

\subsection{Moments}
Information on how the degree is distributed among the nodes of a
undirected network can be obtained either by the degree distribution
$P(k)$, or by the calculation of the moments of the distribution.
The $n$-moment of $P(k)$ is defined as:
\begin{equation}
\langle k^{n} \rangle=\sum_{k}k^{n}P(k).
\end{equation}

The first moment $\langle k \rangle$ is the mean degree. At
arbitrary step $t$, the average vertex degree of $H(q,t)$ is
\begin{equation}
\langle k\label{Kt}
\rangle_t=\frac{2E_t}{N_t}=\frac{2(2q-1)(2q)^{t}}{q(2q)^{t}+3q-2}.
\end{equation}
For large $t$, it is small and approximately equal to a finite value
$4-\frac{2}{q}$.

We can also calculate higher moments of the distribution $P(k)$. For
instance, the second moment, which measures the fluctuations of the
connectivity distribution, is given by
\begin{eqnarray}\label{Ki21}
\langle k^{2}
\rangle_{t}=\frac{1}{N_t}\sum_{t_i=0}^{t}n_v(t_i)\left[k(t_i,t)\right]^{2},
\end{eqnarray}
where $k(t_i,t)$ is the degree of a node at step $t$ which was
generated at step $t_i$. This quality expresses the average of
degree square over all nodes in the network. It has large impact on
the behavior of dynamical processes taking place in networks
\cite{PaVe01a,PaVe01b}.

Substituting Eqs. (\ref{Nv1}), (\ref{Nt1}) and (\ref{Ki1}) into Eq.
(\ref{Ki21}), we derive
\begin{eqnarray}\label{Ki22}
\langle k^{2}
 \rangle_{t}
  & = & \left\{\begin{array}{lc}
{\displaystyle{\frac{4q(2q-1)}
{q+\frac{3q-2}{(2q)^{t}}}\frac{q^{t}-2^{t}}{2^{t}(q^{2}-2q)}+\frac{2(2q-1)q^{t}}{q\cdot2^{t}+\frac{3q-2}{q^{t}}}}
}
& \ \hbox{for}\ q > 2\\
{\displaystyle{\frac{4^{t}(3t+3)}{4^{t}+2}} }
& \ \hbox{for}\  q=2\\
\end{array} \right.\nonumber\\
& \overrightarrow{t\to \infty}& \left\{\begin{array}{lc}
{\displaystyle{\frac{4(q-1)(2q-1)}
{q(q-2)}\left(\frac{q}{2}\right)^{t}} }
& \ \hbox{for}\ q > 2\\
{\displaystyle{3(t+1)} } & \ \hbox{for}\  q=2\\\end{array} \right.
\end{eqnarray}
In this way, second moment of degree distribution $\langle k^{2}
 \rangle$ has been
calculated explicitly, and the result shows that it becomes infinite
for large $t$. In fact, because the degree exponent $\gamma \leq 3$,
all $n$-moments $(n>2)$ diverge. For example, we can analogously get
the third moment as

\begin{equation}\label{Ki3}
\langle k^{3}
 \rangle_{t}=\frac{2^{t+2}q^{3t+3}-2^{t+1}q^{3t+2}+2^{t+3}q^{3t+1}-2^{t+2}q^{3t}-2^{2t+4}q^{t+1}+2^{2t+3}q^{t}}{2^{2t}q^{t+3}-2^{2t+1}q^{t+1}+3\cdot
2^{t}q^{3}-3 q\cdot2^{t+1}-2^{t+1}q^{2}+2^{t+2}}.
\end{equation}

For the special case of $q=2$, it reduces to
\begin{equation}
\langle k^{3}
 \rangle_{q=2,t}=\frac{9\cdot 2^{3t}-6\cdot 2^{2t}}{2^{2t}+2},
\end{equation}
which diverges as an exponential law when $t$ is very large.

\subsection{Clustering Coefficient}
The clustering coefficient defines a measure of the level of
cohesiveness around any given node. By definition, the clustering
coefficient~\cite{WaSt98} $C_i$ of node $i$ is the ratio between the
number of edges $e_i $ that actually exist among the $k_i $
neighbors of node $i$ (i.e. the number of triangles attached to a
vertex $i$) and its maximum possible value, $ k_i( k_i -1)/2 $,
i.e., $ C_i =2e_i/k_i(k_i -1)$. The clustering coefficient of the
whole network is the average of all individual $C_{i}'s$.

Since there are no triangles in the hierarchial lattices, the
clustering coefficient of every node and their average value in
$H(q,t)$ are both zero by definition. However, over the years
generalized clustering coefficients probing higher-order loops have
been proposed \cite{BiCa03,CaPaVe04}. Clearly in these hierarchial
lattices the number of ¡°squares¡± (loops of length 4) is
significantly large, below we will seek to quantify this.

\subsection{Grid Coefficient}
As pointed out above, for hierarchial lattices the usual clustering
coefficient is unable to quantify the order underlying their
structure, which is represented by a grid-like frame, that can be
quantified by evaluating the frequency of rectangular loops (cycles
of length 4). We introduce the \emph{grid coefficient} that allows
us to uncover the presence of a surprising level of triangular grid
ordering in the hierarchial lattices. For simplicity, we call cycles
of length 4 \emph{quadrilaterals}. The grid coefficient
\cite{CaPaVe04} $G_i$ of node $i$ is defined as the ratio of number
of existing quadrilaterals passing by node $i$, $X_i$, to all the
possible number of quadrilaterals attached to node $i$, $Y_i$.

\begin{figure}
\begin{center}
\includegraphics[width=8cm]{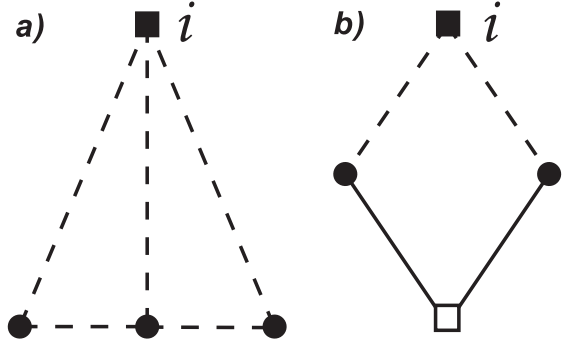}
\caption{\textbf{(a)} An example of a primary quadrilateral, where
    the three outer nodes are directly connected to node
    $i$.  \textbf{(b)} An example of a secondary quadrilateral, where one outer node (empty square) is a second neighbor
    of node $i$.}
\label{grid_coefficient}
\end{center}
\end{figure}
Note that each quadrilateral involving node $i$ is consist of $i$
itself plus three outer nodes, according to whose nature
quadrilaterals can be classified, then the grid coefficient can be
further decomposed into two cases (see Fig.~\ref{grid_coefficient}):
If all the outer nodes are directly attached to $i$, they form a
\textit{primary quadrilateral}; otherwise, if one of the outer nodes
is a second neighbor of $i$, the cycle they form is a
\textit{secondary quadrilateral}. If a node $i$ with degree $k_i$
has $k_{i,2nd}$ second neighbors, then the maximum possible number
of primary quadrilaterals is $Y_i^p = 3 \times (^{k_i}_{3}) =
k_i(k_i-1)(k_i-2)/2$, while the maximum possible number of secondary
quadrilaterals is $Y_i^s = k_{i,2nd} k_i(k_i-1)/2$.  In this way,
for investigating the grid properties of the hierarchial lattices,
one can define three quantities: the primary grid coefficient,
$G_{i}^p=X_i^p/Y_i^p$, the secondary grid coefficient
$G_{i}^s=X_i^s/Y_i^s$, and the total grid coefficient $G_{i} =
(X_i^p +X_i^s)/(Y_i^p+Y_i^s)$, where $X_i^p$ and $X_i^s$ are the
actually existing number of primary and secondary quadrilaterals
involving $i$, respectively. Averaging these quantities over all
nodes, we can obtain the respective average grid coefficients.

There are only secondary quadrilaterals in HLs, so $G_{i}^p=0$. The
analytical expression for the secondary grid coefficient $G_{i}^s$
of the individual node $i$ with degree $k$ can be derived exactly.
In $H(q,t)$, for a node with degree great than two, there is a
one-to-one correspondence between the second neighbors $k_{2nd}$ of
the node and its degree $k$: $k_{2nd}=\frac{k}{q}$. On the other
hand, by construction, for a node with degree $k>2$, for each of its
second neighbor, there are just $(^{q}_{2})$ secondary
quadrilaterals passing by the node and the second neighbor
simultaneously, thus, the existing number of secondary
quadrilaterals is $X^s=\frac{k}{q}(^{q}_{2})$. Therefor, for a node
of degree $k>2$, the exact value of its grid coefficient is
\begin{equation}
G(k)=\frac{q(q-1)}{k(k-1)}.
\end{equation}
So the grid coefficient is a function of degree $k$, following a
power-law behavior of the form $k^{-2}$ for large $k$. It is
interesting to note that a similar scaling has been observed in
several real-life networks \cite{CaPaVe04}.

\subsection{Fractality}
In fact, the hierarchial lattices grow as a inverse renormalization
procedure. To find the fractal dimension, we follow the mathematical
framework proposed in Ref. \cite{SoHaMa06}. By construction, for
large $t$, the different quantities grow as:
\begin{equation}\label{Frac01}
\left\{\begin{array}{lc} {N_t\simeq2q\,N_{t-1} },\\
{k_i(t)=q\,k_i(t-1)},\\
{\mathbb{D}_t=2\,\mathbb{D}_{(t-1)}.}\end{array} \right.
\end{equation}
The first equation is analogous to the multiplicative process
naturally found in many population growth systems. The second
relation denotes the preferential attachment mechanism
\cite{BaAl99}, which yields the power law degree distribution. The
third equation describes the change of the diameter $\mathbb{D}_t$
of the hierarchial lattices $H(q,t)$, where $\mathbb{D}_t$ is
defined as by the longest shortest path between all pairs of nodes
in $H(q,t)$.

From the relations given by Eq. (\ref{Frac01}), we know that these
quantities $N_t$, $k_i(t)$ and $\mathbb{D}_t$ increase by a factor
of $2q$, $q$ and $2$, respectively. Then between any two times
$t_{1}$, $t_{2}$ ($t_{1} < t_{2}$), we can easily obtain the
following relation:
\begin{equation}\label{Frac02}
\left\{\begin{array}{lc} {\mathbb{D}_{t_{2}}=2^{t_{2}-t_{1}}\,\mathbb{D}_{t_{1}}},\\
{N_{t_{2}}=(2q)^{t_{2}-t_{1}}\,N_{t_{1}}},\\
{k_i(t_{2})=q^{t_{2}-t_{1}}\,k_i(t_{1})}.\end{array} \right.
\end{equation}
From Eq. (\ref{Frac02}), we can derive the scaling exponents in
terms of the microscopic parameters: the fractal dimension is $d_B
=\frac{\ln (2q)}{\ln 2}$, and the degree exponent of boxes is $d_k
=\frac{ \ln q}{\ln 2}$. The exponent of the degree distribution
satisfies $\gamma =1+\frac{d_B}{d_k}= 2+\frac{\ln 2}{\ln q}$, giving
the same $\gamma$ as that obtained in the direct calculation of the
degree distribution, see Eq. (\ref{gamma1}).

Note that in a class of deterministic models called pseudo-fractals,
although the number of their nodes increase exponentially, the
additive growth in the diameter with time implies that these
networks are small world. These models do not capture the fractal
topology found in diverse complex networks~\cite{BaRaVi01, DoGoMe02,
CoFeRa04, ZhRoZh06c, JuKiKa02, RaSoMoOlBa02, RaBa03, AnHeAnSi05,
DoMa05,ZhCoFeRo05,ZhRo05}.

\subsection{Degree correlations}
As the field of complex networks has progressed, degree
correlations~\cite{MsSn02,PaVaVe01,VapaVe02,Newman02,Newman03c,BoPa03,BoPa05}
have been the subject of particular interest, because they can give
rise to some interesting network structure effects. Degree
correlations can be conveniently measured by means of the
conditional probability $P(k'|k)$, being defined as the probability
that a link from a node of degree $k$ points to a node of degree
$k'$. In uncorrelated networks, this conditional probability does
not depend on $k$, it takes the form $P(k'|k) = k' P(k') / \langle
k\rangle$ \cite{BoPa03}.

Although degree correlations are formally characterized by
$P(k'|k)$, the direct evaluation of the conditional probability
$P(k'|k)$ in real-life systems is a very difficult task, and usually
gives extremely noisy results because of their finite size. To
overcome this problem, another interesting quantity related to two
node correlations, called \emph{average nearest-neighbor degree}
(ANND), has been proposed. It is a function of node degree, and is
more convenient and practical in characterizing degree-degree
correlations, defined by \cite{PaVaVe01}
\begin{equation}
  k_{nn}(k) = \sum_{k'} k' P(k'|k).
  \label{knn1}
\end{equation}
If there are no degree correlations, Eq. (\ref{knn1}) gives
$k_{nn}(k)=\langle k^{2}\rangle/\langle k\rangle$, i.e. $k_{nn}(k)$
is independent of $k$. Degree correlations are usually quantified by
reporting the numerical value of the slope of $k_{nn}(k)$ as a
function of node degree $k$.

Degree correlations quantified by ANND have led to a first
classification of complex networks. When $k_{\rm nn}(k)$ increases
with $k$, it means that nodes have a tendency to connect to nodes
with a similar or larger degree. In this case the network is defined
as\emph{ assortative} \cite{Newman02,Newman03c}. In contrast, if
$k_{\rm nn}(k)$ is decreasing with $k$, which implies that nodes of
large degrees are likely to have the nearest neighbors with small
degrees, then the network is said to be \emph{disassortative}.

We can exactly calculate $k_{\rm nn}(k)$ for the hierarchial
lattices. By construction, for nodes with degree greater than 2, the
degrees of their neighbors are 2. Then we have
\begin{equation}
k_{\rm nn}(k>2)=2.
\end{equation}
 For those nodes having degree 2, their average
nearest-neighbor degrees are
\begin{eqnarray}
k_{\rm nn}(2)&=&{1\over 2 L_v(t)}
  \Bigg(\sum_{t'_i=0}^{t'_i=t-1} L_v(t'_i) \left[k(t'_i,t)\right]^{2}\Bigg)\nonumber\\
  &=&\left\{\begin{array}{lc} {\displaystyle{2\,t} }
& \ \hbox{for}\ q=2,\\
{\displaystyle{\frac{2q}{q-2} \left[ \left(\frac{q}{2}\right)^{t}-1
\right ]}}
& \ \hbox{for}\  q>2.\\
\end{array} \right.
\end{eqnarray}
Thus $k_{\rm nn}(2)$ grows linearly or exponentially with time for
$q=2$ and $q>2$, respectively. As the nodes with degree 2 are only
connected to higher degree nodes, $k_{\rm nn}(2)$ is significantly
high.

Degree correlations can also be described by a Pearson correlation
coefficient $r$ of degrees at either end of a link. It is defined as
\cite{Newman02,Newman03c,DoMa05,RaDoPa04}
\begin{equation}\label{Pearson}
r={\langle k\rangle\langle k^2 k_{\rm nn}(k)\rangle -
    \langle k^2\rangle^2 \over
    \langle k\rangle \langle k^3\rangle - \langle k^2\rangle^2}.
\end{equation}
If the network is uncorrelated, the correlation coefficient equals
zero. Disassortative networks have $r<0$, while assortative graphs
have a value of $r>0$. Substituting Eqs. (\ref{Ki1}), (\ref{Ki22})
and (\ref{Ki3}) into Eq. (\ref{Pearson}), we can easily see that for
$t>1$, $r$ of $H(q,t)$ is always negative, indicating
disassortativity.

Disassortative features in protein interaction networks were found
and explained by Maslov and Sneppen \cite{MsSn02} on the level of
interacting proteins and genetic regulatory interactions. According
to their results links between highly connected nodes are
systematically suppressed, while those between highly connected and
low-connected pairs of proteins are favored.

\subsection{Average path length}

Shortest paths play an important role both in the transport and
communication within a network and in the characterization of the
internal structure of the network. We represent all the shortest
path lengths of $H(q,t)$ as a matrix in which the entry $d_{ij}$ is
the geodesic path from node $i$ to node $j$, where geodesic path is
one of the paths connecting two nodes with minimum length. The
maximum value of $d_{ij}$ is called the diameter of the network. A
measure of the typical separation between two nodes in the
hierarchical lattices is given by the average path length
$\bar{d}_{t}$, also known as characteristic path length, defined as
the mean of geodesic lengths over all couples of nodes at the $t$th
level.

For general $q$,  it is not easy to derive a closed formula for the
average path length $\bar{d}_{t}$ of $H(q,t)$. However, in the
Appendix, we have obtained exact analytic expressions for
$\bar{d}_{t}$ of $H(3,t)$, while the exact value of $\bar{d}_{t}$ of
$H(2,t)$ has been obviously obtained in Ref. \cite{HiBe06}. For
$q=3$ we find

\begin{eqnarray}\label{eq:6}
\bar{d}_{t} = \frac{1243\cdot2^t -475 \cdot2^{t+2}3^t+275
\cdot2^{t+3}3^t+275 \cdot2^t3^{2t+1} +19 \cdot2^{3t+2}3^{2t+1}-11
\cdot3^{t+2}4^{t+1}}{44(2+2^t3^{t+1})(7+2^t3^{t+1})}\nonumber \\
\overrightarrow{t\to \infty} \frac{19}{33}2^t\,,
\end{eqnarray}
leading to an exponential growth in the APL. Since in this case,
$N_t \sim 6^t$ for large $t$, we have $\bar{d}_{t} \sim
N_t^{\log_{6}2}$. While in another special case $q=2$, $\bar{d}_{t}
\sim N_t^{1/2}$ \cite{HiBe06}. So for small $q$, the hierarchical
lattices are not small worlds. We conjecture that for $H(q,t)$,
their APL scales as $\bar{d}_{t} \sim N_t^{\log_{(2q)}2}$, which is
similar to that of a hypercubic lattice of dimension
$\frac{1}{\log_{(2q)}2}$. Low-dimensional regular lattices do not
show the small-world behavior of typical node-node distances. It is
straightforward to show that for a regular lattice in $D$ dimensions
which has the shape of a square or (hyper)cube of side $l$, and
therefore has $N=l^{D}$ nodes, the APL increases as $l$, or
equivalently as $N^{1/D}$ \cite{Ne00}.

So we have shown that $\bar{d}_{t}$ of $H(q,t)$ has the power-law
scaling behavior of the number of nodes $N_{t}$. It is not hard to
understand. As an example, let us look at the scheme of the growth
of a particular case $q=2$. Each next step in the growth of $H(2,t)$
doubles the APL between a fixed pair of nodes. The total numbers of
nodes and edges increase four-fold (asymptotically, in the infinite
limit of $t$), see Eq. (\ref{Nt1}). Thus the APL $\bar{d}_{t}$ of
$H(2,t)$ grows as a square power of the node number in $H(2,t)$.

\section{Small-world hierarchical lattices}
In this section, we will discuss the construction and properties of
small-world hierarchical lattices (SWHLs). Our goal is to reduce the
diameter enough so as to get a logarithmically growing diameter,
while maintaining the original structure of hierarchical lattices
studied in preceding section. All these can be attained by adding a
new central point and connecting it with a certain set of original
nodes, which is akin to the ideas presented in Refs.
\cite{DoMe00,NiMoLaHo02,BaFeDa06}.

\begin{figure}
\begin{center}
\includegraphics[width=6cm]{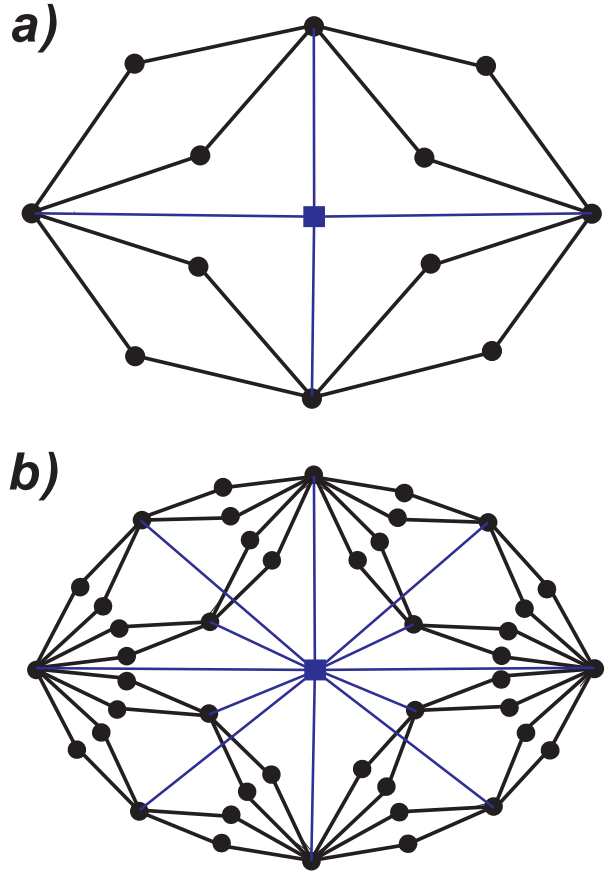}
\caption{(Color online) Construction of the small-world hierarchical
lattices, with (a), (b) denoting $SH_{1}(2,2)$ and $SH_{2}(2,3)$,
respectively.} \label{fig4}
\end{center}
\end{figure}

We denote the small-world hierarchical lattices as  $SH_{m}(q,t)$.
From the previous section, we can know that for every $t\geq 2$ and
$m =1, 2, 3, \cdots , t-1$, $H(q,t)$ can be seen as $(2q)^{t-m}$
copies of $H(q,m)$, with some node identifications. $SH_{m}(q,t)$
($1 \leq m \leq t-1$) is the graph obtained by joining a new node to
every hub (node of highest degree) of every copy of $H(q,m)$, see
Fig. \ref{fig4}. In other words, the new node is connected to those
old nodes introduced at step $t-m$ or earlier, thus the number of
new edges is exactly the number of vertices of $H(q,t-m)$. For
$m=0$, the new graph is out of the scope of $SH_{m}(q,t)$. In this
case, for simplicity, we also denote the new network $SH_{m}(q,t)$,
where the central node connects to all the nodes in $H(q,t)$.

The diameter of this new graph depends on the value of $m$. We will
show that, for some of the values of $m$, the diameter of
$SH_{m}(q,t)$ exhibits a slow (logarithmic) increase with the total
number of network nodes. Thus, this construction gives us
small-world graphs. Next we give the properties of small-world
hierarchical lattices.

The order (number of all nodes) of $SH_{m}(q,t)$ is one plus the
order of $H(q,t)$. The size (number of all edges) of $SH_{m}(q,t)$
is the size of $H(q,t)$, plus the number of added edges. Since the
number of added edges is the order of $H(q,t-m)$, according to
Eqs.~(\ref{Nt1}) and~(\ref{Et1}), we can easily see that the order
and size of $SH_{m}(q,t)$ is $N_{m,t}=\frac{q(2q)^{t}+5q-3}{2q-1}$
and $E_{m,t}=\frac{(2q)^{t+1}-(2q)^{t}+q(2q)^{t-m}+3q-2}{2q-1}$,
respectively.

Because the addition of the new node has little effect on the degree
distribution, $SH_{m}(q,t)$ also follow power law degree
distribution with the same degree exponent $\gamma$ as $H(q,t)$.
Additionally, for any $m\geq 1$, $SH_{m}(q,t)$ have no triangles,
the clustering coefficient is zero as their counterparts $H(q,t)$.
Analogously, $SH_{m}(q,t)$ are self-similar with the identical
fractal dimension $d_{B}$ as $H(q,t)$ \cite{SoHaMa06,BaFeDa06}.

\begin{figure}
\begin{center}
\includegraphics[width=8cm]{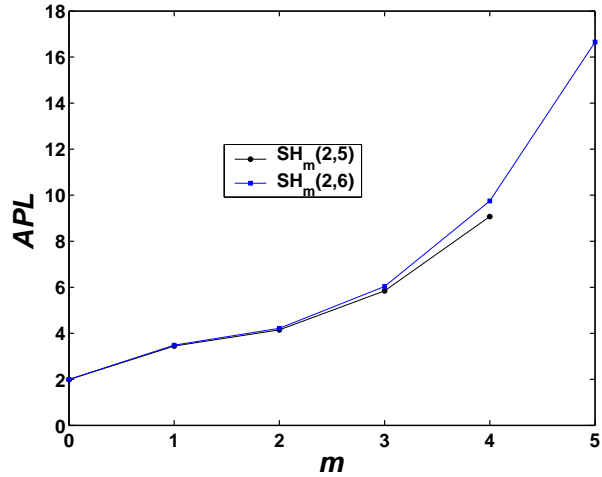}
\caption{(Color online) The dependence of average path length (APL)
of $SH_{m}(2,5)$ and $SH_{m}(2,6)$ on $m$. One can see that APL
increases very quickly as $m$ grows.} \label{fig5}
\end{center}
\end{figure}

Different from $H(q,t)$, $SH_{m}(q,t)$ have small-world property.
For the sake of convenient expression, let us denote by
$Diam[H(q,t)]$ the diameter of $H(q,t)$ and by $Diam[SH_{m}(q,t)]$
the diameter of $SH_{m}(q,t)$. Obviously, $Diam[H(q,t)]=2^{t}$. To
compute $Diam[SH_{m}(q,t)]$ we need only observe that, in $H(q,m)$,
every node is at distance at most $\frac{Diam[H(q,m)]}{2}=2^{m-1}$
from the set of vertices of the hubs. An upper bound for
$Diam[SH_{m}(q,t)]$ is $Diam[H(q,m)]+2=2^{m}+2$. It can easily be
seen that this is also a lower bound. Therefore,
$Diam[SH_{m}(q,t)]=2^{m}+2$. Since the order $N_{m,t}$ of
$SH_{m}(q,t)$ is $\frac{q(2q)^{t}+5q-3}{2q-1}$, if $m\leq
\log_{2}t$, then $Diam[SH_{m}(q,t)]\leq t+2$ is small and scales
logarithmically with the number of network nodes. Here we do not
give the exact expression for the average path length (APL) of
$SH_{m}(q,t)$, instead in Fig.~\ref{fig5} we present APL of
$SH_{m}(q,t)$ as a function of $m$. It is shown that APL becomes
larger as $m$ is increased.

To summarize, in the section, we proposed a construction of
small-world hierarchical lattices. In the construction of these
small-world lattices, the underlying structure of the original
lattices is preserved. We have shown that all of these new graphs
are fractal and have a logarithmic diameter.

\section{Relative robustness to international attacks}
As discussed in previous section, close to many real-life networks,
the hierarchical lattices are simultaneously self-similar and
scale-free. Therefore, it is worthwhile to investigate the processes
taking place upon them and directly compare these results with just
scale-free networks (like BA networks). These comparisons may give
us deep insight into the dynamic properties of networks. In the
following we will investigate intentional damage (attack) and
synchronization, respectively. This section is devoted to the
robustness, while next section is concerned with collective
synchronization behavior.

Robustness refers to the ability of a network to avoid
malfunctioning when a fraction of its constituents is damaged. This
is a topic of obvious practical reasons, as it affects directly the
efficiency of any process running on top of the network, and it is
one of the first issues to be explored in the literature on complex
networks \cite{AlJeBa00}. Here we shall focus only on the
topological aspects (especially self-similarity) of robustness,
caused by targeted removal of nodes, because there is a strong
correlation between robustness and network topology \cite{AlJeBa00,
CaNeStWa00,CoErAvHa01,DoMe01,HoKiYoHa02}.

One of the most important measures of the robustness of a network is
its integrity, which is characterized by the presence of its giant
connected component \cite{AlJeBa00}. We call a network robust if it
contains a giant cluster comprised of most of the nodes even after a
fraction of its nodes are removed. Then, to know network robustness,
first of all, one must study the variation of the giant component.

For the study of attack vulnerability of the network, the selection
procedure of the order in which vertices are removed is an open
choice \cite{HoKiYoHa02}. One may of course maximize the destructive
effect at any fixed number of removed vertices. However, this
requires the knowledge of the whole network structure, and
pinpointing the vertex to attack in this way makes a very
time-demanding computation. A more tractable choice is to select the
vertices in the descending order of degrees in the initial network
and then to remove vertices one by one starting from the vertex with
the highest degree; this attack strategy will be used in the present
paper.

\begin{figure}
\begin{center}
\includegraphics[width=9cm]{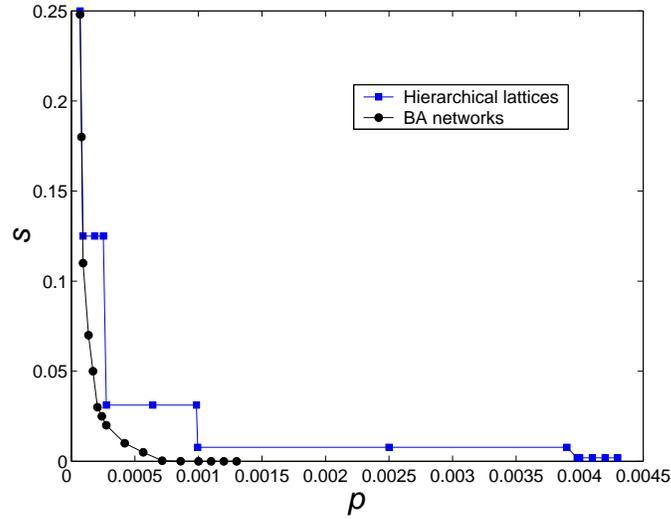}
\caption{(Color online) Vulnerability under intentional attack of a
BA network with average degree 4 and a $H(2,8)$. Both networks have
the same degree exponent ($\gamma=3$), the same number of nodes
(43,692), and their clustering coefficient is small. Moreover, by
construction, $SH_{m}(2,8)$ have similar robustness as $H(2,8)$,
which rule out the effect of average path length on targeted attack.
Thus, the difference in the resilience seen in this figure is
attributed to fractality and the different degree of
anticorrelations.} \label{fig6}
\end{center}
\end{figure}

Figure~\ref{fig6} shows the performances of BA and hierarchial
lattices under intentional attack. We plot the relative size of the
largest cluster, $s$, after removing a fraction $p$ of the largest
hubs for both networks. One can find that the non-fractal scale-free
BA networks are more sensitive to sabotage of a small fraction of
the nodes, leading support to the view of Song et. al.
\cite{SoHaMa06}. While both networks collapse at a finite fraction
$p_c$, evidenced by the decrease of $s$ toward zero, the fractal
network has a significantly larger threshold ($p_c\approx 0.004$)
compared to the non-fractal threshold ($p_c\approx 0.001$), the
former threshold is about 4 times than the latter, suggesting a
significantly higher robustness of the fractal networks to
intentional attacks. Also, it is interesting to note that for the
hierarchical lattices, $s$ is a function of $p$ with a
staircase-like form.

It is not strange at all that a giant connected component in
self-similar networks is robust against the targeted deletion of
nodes, while non-fractal scale-free networks are extremely
vulnerable to targeted attacks on the hub. In non-fractal
topologies, the hubs are connected and form a central compact core,
such that the removal of a few of the largest hubs has catastrophic
consequences for the network. In self-similar networks, hubs are
more dispersed (see Fig.~\ref{fig2}), their disassortativity and
self-similar property significantly increases the robustness against
targeted attacks. This could explain why some real-life networks
have evolved into a fractal and disassortative architecture
\cite{SoHaMa05,SoHaMa06}.

\section{Synchronization}

The ultimate goal of the study of network structure is to study and
understand the workings of systems built upon those networks.
Recently, along with the study of purely structural and evolutionary
properties~\cite{AlBa02, DoMe02}, there has been increasing interest
in the interplay between the dynamics and the structure of complex
networks~\cite{Ne03,BoLaMoChHw06}. One particular issue attracting
much attention is the synchronizability of oscillator coupling
networks \cite{St03}. Synchronization is observed in diverse natural
and man-made systems and is directly related to many specific
problems in a variety of different disciplines. It has found
practical applications in many fields including communications,
optics, neural networks and geophysics
\cite{PeCa90,WiRa90,HaSo92,CuOp93,Vi99,Ot00}. After studying the
relevant characteristics of network structure, which is described in
the previous sections, we will study the synchronization behavior on
the networks.

We follow the general framework proposed in \cite{BaPe02,PeBa98},
where a criterion based on spectral techniques was established to
determine the stability of synchronized states on networks. Consider
a network of $N$ identical dynamical systems with linearly and
symmetric coupling between oscillators. The set of equations of
motion for the system are
\begin{equation}
\dot{\textbf{x}}_i=\textbf{F}(\textbf{x}_i)+\sigma\sum_{j=1}^NG_{ij}\textbf{H}(\textbf{x}_j),
\end{equation}
where $\dot{\textbf{x}}_i=\textbf{F}(\textbf{x}_i)$ governs the
dynamics of each individual node, $\textbf{H}(\textbf{x}_j)$ is the
output function and $\sigma$ the coupling strength, and $G_{ij}$ is
the Laplacian matrix, defined by $G_{ii}=k_i$ if the degree of node
$i$ is $k_i$, $G_{ij}=-1$ if nodes $i$ and $j$ are connected, and
$G_{ij}=0$ otherwise.

Since matrix $G$ is positive semidefinite and each rows of it has
zero sum, all eigenvalues of $G$ are real and non-negative and the
smallest one is always equal to zero. We order the eigenvalues as
$0=\lambda_1\leq\lambda_2\leq\cdots\leq\lambda_{N}$. Then one can
use the ratio of the maximum eigenvalue $\lambda_{N}$ to the
smallest nonzero one $\lambda_2$ to measure the synchronizability of
the network \cite{BaPe02,PeBa98}. If the eigenratio
$R=\lambda_{N}/\lambda_2$ satisfies $R<\alpha_2/\alpha_1$, we say
the network is synchronizable. Here the eigenratio $R$ depends on
the the network topology, while $\alpha_2/\alpha_1$ depends
exclusively on the dynamics of individual oscillator and the output
function. Ratio $R=\lambda_{N}/\lambda_2$ represents the
synchronizability of the network: the larger the ratio, the more
difficult it is to synchronize the oscillators, and vice versa.

\begin{figure}
\begin{center}
\includegraphics[width=8cm]{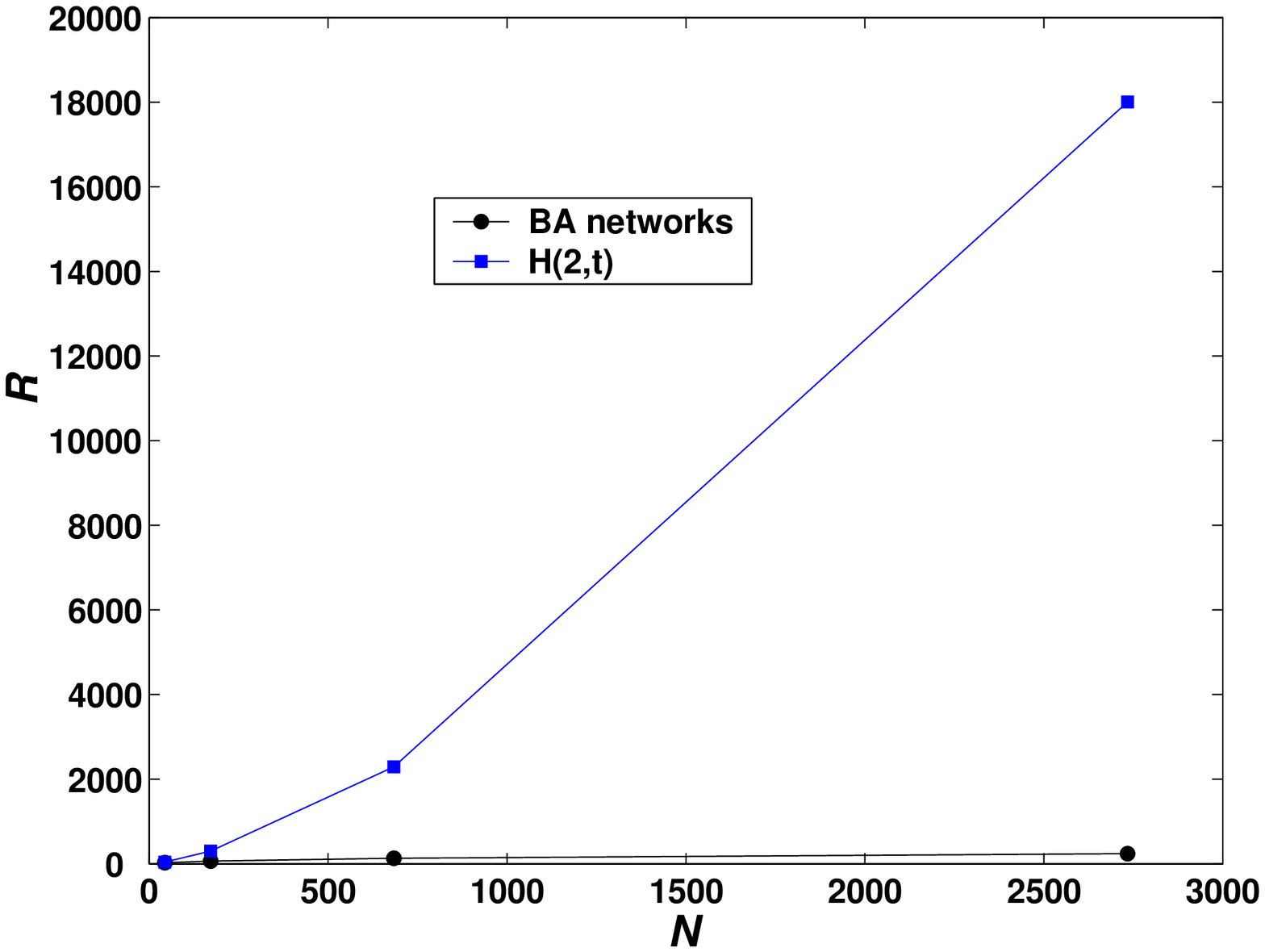}\
\includegraphics[width=8cm]{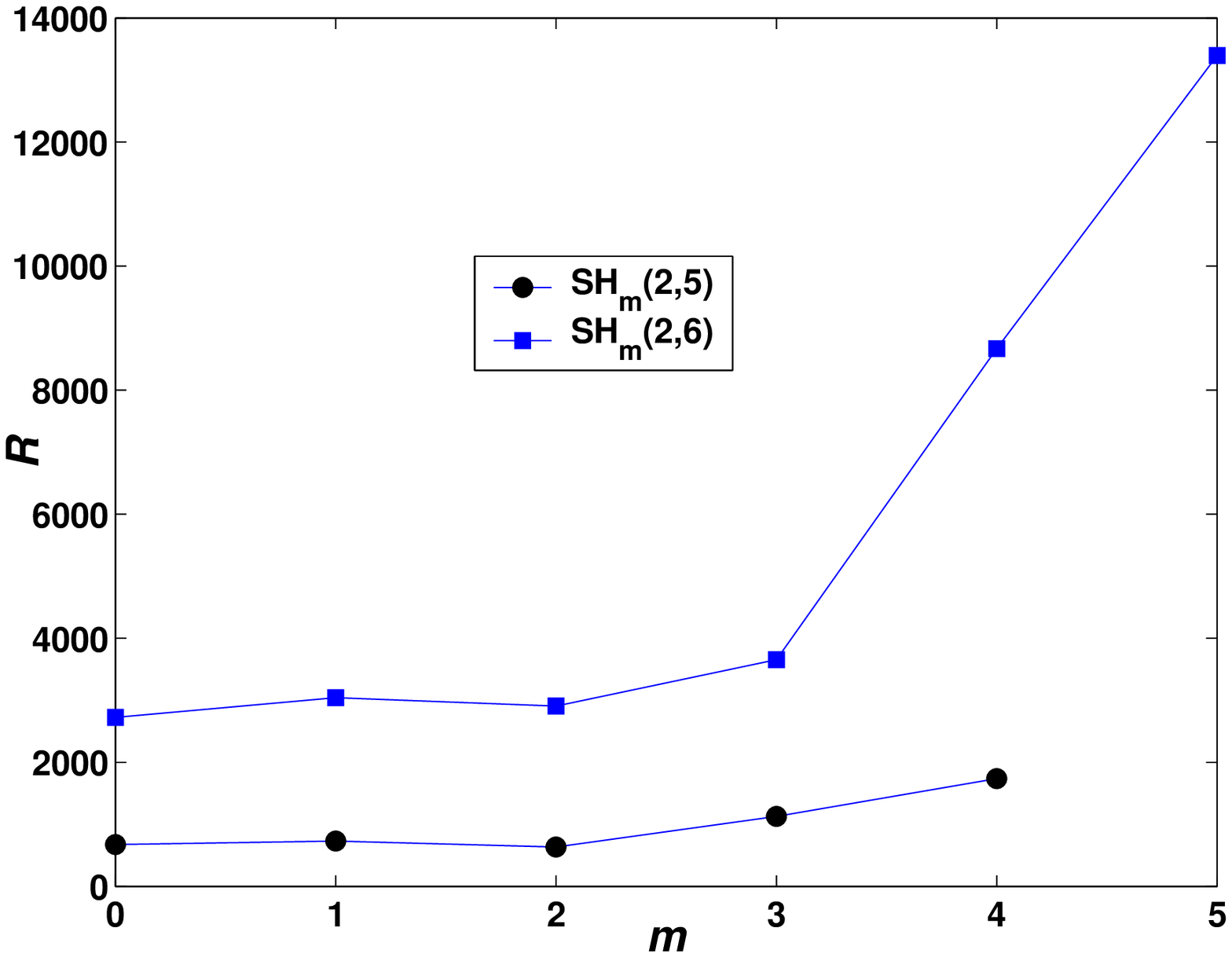}\
\caption{(Color online) (Top) The eigenratio $R$ as a function of
network order $N$ for BA networks with average degree 4 and
$H(2,t)$. All quantities for BA networks are averaged over 50
realizations. (Bottom) The dependence of eigenratio $R$ of
$SH_{m}(2,5)$ and $SH_{m}(2,6)$ on the value of $m$.} \label{fig7}
\end{center}
\end{figure}

After reducing the issue of synchronizability to finding eigenvalues
of the Laplacian matrix $G$, we now investigate the synchronization
of our networks. Figure~\ref{fig7} shows the synchronizabily of
$H(2,t)$, $SH_{m}(2,5)$, $SH_{m}(2,6)$, as well as the BA networks.
One can see that for the same network order, $R$ of the BA networks
is much smaller than that of hierarchical lattices $H(2,t)$, which
implies that the synchronizability of the former is much better.
While for $SH_{m}(2,5)$ and $SH_{m}(2,6)$, the dependence relation
of eigenratio $R(m)$ on $m$ is more complicated: $R(0)< R(2) <R(1) <
R(x|x>2)$; for $m>2$, $R(m)$ increases with $m>2$.

Why coupling systems on the BA networks, $H(2,t)$ and $SH_{m}(2,t)$
exhibit very different synchronizability? Previously reported
results have indicated that underlying network structures play
significant roles in the synchronizability of coupled oscillators.
However, the key structural feature that determines the collective
synchronization behavior remains unclear. Many works have discussed
this issue. Some authors believe that shorter APL tends to enhance
synchronization \cite{BaPe02,GaHu00,ZhZhWa05}. In contrast,
Nishikawa\emph{ et al.} reported that synchronizability is
suppressed as the degree distribution becomes more heterogeneous,
even for shorter APL \cite{NiMoLaHo03}. In Ref.~\cite{GoYaYa05}, the
authors asserted that larger average node degree corresponds to
better synchroizability.

All these may rationally explain the relations between
synchroizability and network structure in some cases, but do not
well account for the difference of synchroizability between the
graphs under consideration: as known from the preceding section, for
fixed $t$, the APL of $SH_{m}(2,t)$ increases with $m$, but $R$ does
not always decrease with $m$; BA networks and $H(2,t)$ have
identical degree exponent $\gamma$, while their synchroizability
differs very much; in addition, the average degree of $SH_{1}(2,t)$
is higher than that of $SH_{2}(2,t)$, but the former is more
difficult to synchronize than the latter. All these show that the
degree distribution is generally not sufficient to characterize the
synchronizability of scale-free networks \cite{AtBiJo06,HaSwSc06},
and that smaller average path length does not necessarily predict
better synchroizability \cite{HaSwSc06}. We speculate that the
synchroizability on BA networks is better than on $H(2,t)$ and
$SH_{m}(2,t)$ rests mainly with the self-similar structure. The
genuine reasons need further research.

\section{Conclusions}
In conclusion, we have studied a family of deterministic networks
called hierarchical lattices (HLs) from the perspective of complex
networks. The deterministic self-similar construction allow one to
derive analytic exact expressions for the relevant features of HLs.
Our results shows that HLs exhibit many interesting topological
properties: they follow a power-law degree distribution with
exponent tuned from 2 to 3; their clustering coefficient is null but
their grid coefficient follows a power-law phenomenon; they are not
small-world, the APL scales like a power-law in the number of nodes;
they have a fractal topology with a general fractal dimension; and
they are disassortative networks. Our results indicate it is not
true that a power-law degree alone create small-world networks
\cite{KlEg02}, and further support the conjecture that scale-free
networks with fractal scaling are disassortative. The
disassortativity property is ease to understand by checking the
growth process of HLs, where the rich (large nodes) get richer but
at the expense of the poor (small nodes). In other words, the hubs
prefer to grow by connections with less-connected nodes rather to
other hubs, which leads to disassortativity. So we have found a good
example---hierarchical lattices---which show that self-similar
scale-free networks are preferably disassortative in their
degree-degree correlations.

We have also introduced a deterministic construction of a family of
small-world hierarchical lattices (SWHLs) and investigated their
topological characteristics. We have shown that the some basic
structure features of the hierarchical lattices (HLs) are preserved,
including degree distribution, clustering coefficient and
fractality, while the small-world phenomenon arises.

In addition, we have studied the dynamical processes such as
intention damage and collective synchronization and have shown the
comparisons between HLs and BA networks as well as SWHLs. We have
found that self-similarity and disassortativity increase the
robustness of networks under intentional attacks. Some qualitative
explanations have been given showing that a fractal and
disassortative topology structure is more robust. Although HLs and
SWHLs have relative better robustness, they exhibit poorer
synchronizability than BA networks based possibly on the same
reasons as that of their insensitiveness to sabotage. We have shown
that sychronizability of scale-free networks is not an intrinsic
property of the exponent of degree distribution, and that small APL
does not imply good synchronizability.

In future, it would be worth searching for other stochastic networks
displaying finite fractal dimension spectra. Moreover, it is more
interesting that the presence of self-similarity and
disassortativity, as well as the absence of small-world properties
of the HLs might have a relevant effect on the other dynamic process
such as epidemic spreading~\cite{ZhZhZh06}, routing traffic~\cite
{ZhLaPaYe05,WaYiYaWa06}, games~\cite{WaReChWa06} and
transport~\cite{ZhXiCh98} taking place on the networks.

\subsection*{Acknowledgment}
This research was supported by the National Natural Science
Foundation of China under Grant Nos. 60373019, 60573183, and
90612007.

\appendix

\section{Derivation of the average path length for $q=3$}

\begin{figure}
\begin{center}
\includegraphics[width=8cm]{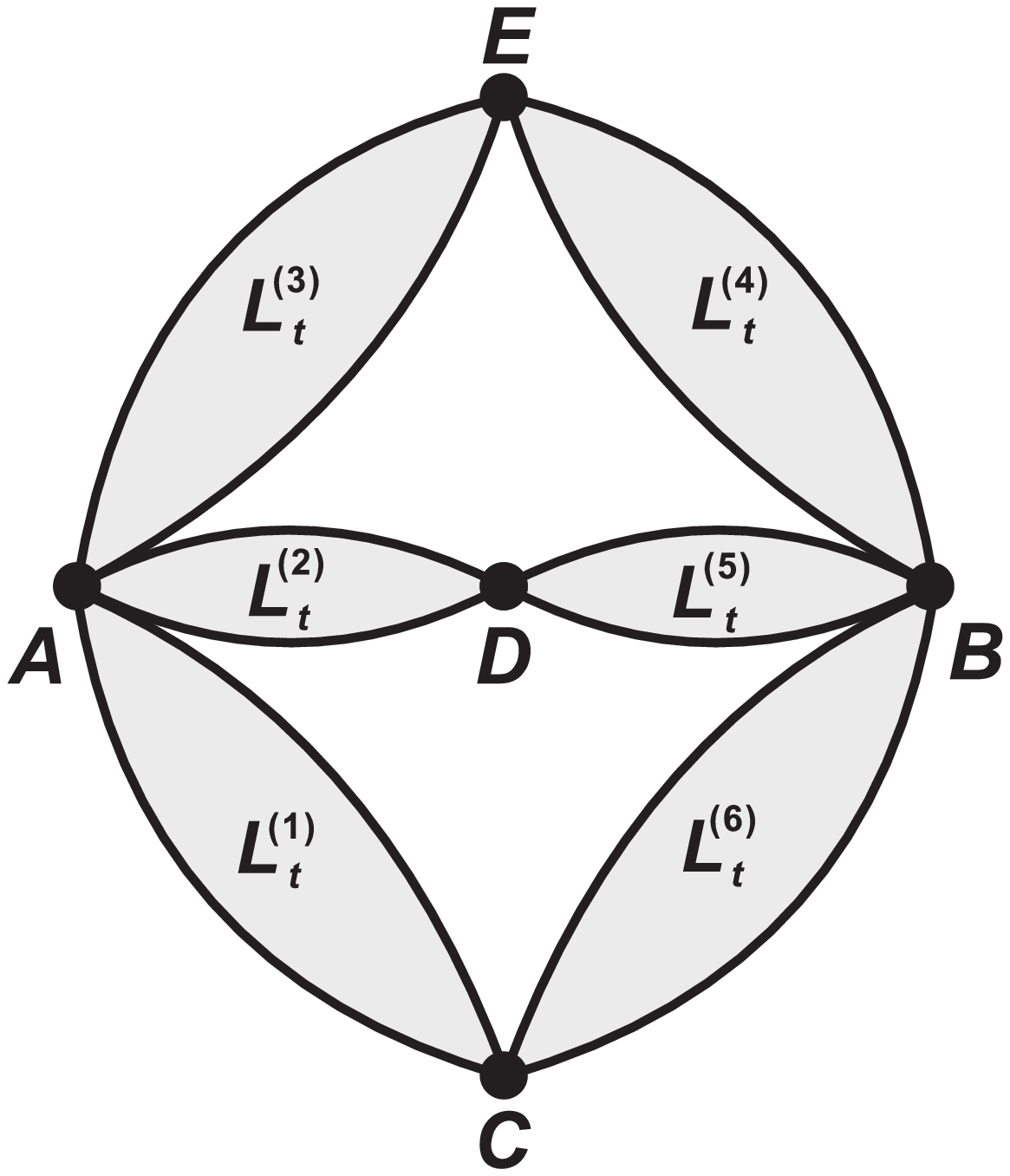}
\caption{For $q=3$, the hierarchical lattice after $t+1$
    construction steps, $L_{t+1}$, is composed of six copies of
    $L_t$ denoted as
    $L_t^{(\chi)}$ $(\chi=1,2,\cdots,6)$, which are
    connected to one another as above. } \label
{apfig2}
\end{center}
\end{figure}

We denote the set of nodes constituting the hierarchical lattices
$H(q,t)$ after $t$ construction steps as $L_{q,t}$.  Then the APL
for $L_{q,t}$ is defined as:
\begin{equation}\label{eq:app4}
  \bar{d}_{t}  = \frac{D_t}{N_t(N_t-1)/2}\,,
\end{equation}
where
\begin{equation}\label{eq:app5}
  D_t = \sum_{i\neq j,\,\, i \in L_{q,t},\, j \in L_{q,t}} d_{ij}
\end{equation}
denotes the sum of the chemical distances between two nodes over all
pairs, and $d_{ij}$ is the chemical distance between nodes $i$ and
$j$. Although there are some difficulties in obtaining a closed
formula for $\bar{d}_{t}$ holding true for all $q$, the hierarchical
lattices have a self-similar structure that allows one to calculate
$\bar{d}_{t}$ analytically according to different $q$.  As shown in
Fig.~\ref{apfig2}, the lattice $L_{q,t+1}$ may be obtained by
joining $2q$ copies of $L_{q,t}$ at the hubs, which are labeled as
$L_{q,t}^{(\alpha)}$, $\alpha=1,2,\cdots,2q$.  Then we can write the
sum $D_{t+1}$ as
\begin{equation}\label{eq:app6}
  D_{t+1} = 2q\,D_t + \Delta_t\,,
\end{equation}
where $\Delta_t$ is the sum over all shortest paths whose endpoints
are not in the same $L_{q,t}$ branch. The solution of
Eq.~\ref{eq:app6} is
\begin{equation}\label{eq:app8}
  D_t = (2q)^{t-1} D_1 + \sum_{m=1}^{t-1} (2q)^{t-m-1} \Delta_m\,.
\end{equation}
The paths that contribute to $\Delta_t$ must all go through at least
one of the $q+2$ edge nodes (for $q=3$ see Fig.~\ref{apfig2}, where
$\textbf{\emph{A}}$, $\textbf{\emph{B}}$, $\textbf{\emph{C}}$,
$\textbf{\emph{D}}$, $\textbf{\emph{E}}$ are the five edge nodes) at
which the different $L_{q,t}$ branches are connected. The analytical
expression for $\Delta_t$ for general $q$, called the crossing
paths, is not easy to derive. We trace the formula only for the
particular case of $q=3$ as follow.

In what follows, we write $L_{3,t}$ as $L_{t}$ for brevity. Denote
$\Delta_t^{\alpha,\beta}$ as the sum of all shortest paths with
endpoints in $L_t^{(\alpha)}$ and $L_t^{(\beta)}$. If
$L_t^{(\alpha)}$ and $L_t^{(\beta)}$ meet at an edge node,
$\Delta_t^{\alpha,\beta}$ rules out the paths where either endpoint
is that shared edge node.  If $L_t^{(\alpha)}$ and $L_t^{(\beta)}$
do not meet, $\Delta_t^{\alpha,\beta}$ excludes the paths where
either endpoint is any edge node.  Then the total sum $\Delta_t$ is
\begin{eqnarray}
\Delta_t =& \,\Delta_t^{1,2} + \Delta_t^{1,3} + \Delta_t^{1,4}+\Delta_t^{1,5}+ \Delta_t^{1,6}+ \Delta_t^{2,3}\nonumber\\
&+ \Delta_t^{2,4}+\Delta_t^{2,5}+ \Delta_t^{2,6}+ \Delta_t^{3,4}+ \Delta_t^{3,5}+\Delta_t^{3,6}\nonumber\\
&+ \Delta_t^{4,5}+ \Delta_t^{4,6}+\Delta_t^{5,6}-5\cdot 2^{t+1}\,.
\label{eq:app10}
\end{eqnarray}
The last term at the end compensates for the overcounting of certain
paths: the shortest path between $\textbf{\emph{A}}$ and
$\textbf{\emph{B}}$, with length $2^{t+1}$, is included in
$\Delta_t^{1,6}$, $\Delta_t^{2,5}$ and $\Delta_t^{3,4}$; the
shortest path between $\textbf{\emph{C}}$ and $\textbf{\emph{E}}$,
with length $2^{t+1}$, is included in both $\Delta_t^{1,3}$ and
$\Delta_t^{4,6}$; the shortest path between $\textbf{\emph{D}}$ and
$\textbf{\emph{E}}$, with length $2^{t+1}$, is included in both
$\Delta_t^{2,3}$ and $\Delta_t^{4,5}$; the shortest path between
$\textbf{\emph{C}}$ and $\textbf{\emph{D}}$, also with length
$2^{t+1}$, is included in both $\Delta_t^{1,2}$ and
$\Delta_t^{5,6}$.

By symmetry, $\Delta_t^{1,2} = \Delta_t^{1,3} = \Delta_t^{2,3} =
\Delta_t^{5,6} = \Delta_t^{4,5} = \Delta_t^{4,6} = \Delta_t ^{1,6} =
\Delta_t^{2,5} = \Delta_t^{3,4}$ and $\Delta_t^{1,4} =
\Delta_t^{1,5} = \Delta_t^{2,4} = \Delta_t^{2,6} = \Delta_t^{3,5} =
\Delta_t^{3,6}$, so that
\begin{equation}\label{eq:app11}
\Delta_t = 9 \Delta_t^{1,2} + 6\Delta_t^{1,4} - 5\cdot 2^{t+1}\,,
\end{equation}
where $\Delta_t^{1,2}$ is given by the sum

\begin{eqnarray}
  \Delta_t^{1,2} &=& \sum_{i \in L_t^{(1)},\,\,j \in L_t^{(2)}, \, i\ne A,\,j \ne A} d_{ij} \nonumber\\
  &=& \sum_{i \in L_t^{(1)},\,\,j\in L_t^{(2)},\, i\ne A,\,j \ne A} (d_{iA} + d_{Aj}) \nonumber\\
  &=& (N_t-1)\sum_{i \in L_t^{(1)}} d_{iA} + (N_t-1) \sum_{j
    \in L_t^{(2)}} d_{Aj} \nonumber\\
  &=& 2(N_t-1)\sum_{i \in L_t^{(1)}} d_{iA}\,,
\label{eq:app12}
\end{eqnarray}
where $\sum_{i \in L_t^{(1)}} d_{iA} = \sum_{j \in L_t^{(2)}}
d_{Aj}$ have been used.  To find $\sum_{i \in
  L_t^{(1)}} d_{iA}$, we examine the structure of the
hierarchical lattice at the $t$th level.  In $L_t^{(1)}$, there is
$\nu_t(m)$ points with $d_{iA} = m$, where $1 \le m \le 2^t$, and
$\nu_t(m)$ can be written recursively as

\begin{equation}\label{eq:app13}
\nu_t(m)= \left\{\begin{array}{lc} {\displaystyle{3^t} }
& \ \hbox{if $m$ is odd} \ \\
{\displaystyle{\nu_{t-1}(\frac{m}{2})} } & \ \hbox{if $m$ is even}\
\\\end{array} \right.
\end{equation}

We can write $\sum_{i \in
  L_t^{(1)}} d_{iA}$ in terms of $\nu_t(m)$ as
\begin{equation}\label{eq:app14}
f_t \equiv \sum_{i \in L_t^{(1)}} d_{iA} = \sum_{m=1}^{2^t} m\cdot
\nu_t(m)\,.
\end{equation}
Eqs.~(\ref{eq:app13}) and (\ref{eq:app14}) relate $f_t$ and
$f_{t-1}$, which allow one to resolve $f_t$ by induction as follow:
\begin{eqnarray}
f_t &= \sum_{k=1}^{2^{t-1}} (2k-1) 3^t + \sum_{k=1}^{2^{t-1}}
2k\cdot
\nu_{t-1}(k)\nonumber\\
&= 3^t2^{2t-2} + 2f_{t-1}\nonumber\\
&= \frac{1}{5}2^{t-2} (14 + 6^{t+1})\,, \label{eq:app15}
\end{eqnarray}
where $f_1 = \nu_1(1) + 2\nu_1(2)= 5$ has been used.  Substituting
Eq.~(\ref{eq:app15}) and $N_t = \frac{7+3\times6^t}{5} $ into
Eq.~(\ref{eq:app12}), we obtain
\begin{equation}\label{eq:app16}
\Delta_t^{1,2} = \frac{1}{25}2^{t-1}(3\times6^t+2)(6^{t+1}+14)\,.
\end{equation}

Continue analogously,
\begin{eqnarray}
\Delta_t^{1,4} =& \sum_{i \in L_t^{(1)},\,\,j\in L_t^{(4)}, \,i\ne A, C,\,\,j\ne B, E} d_{ij}\nonumber\\
=& \sum_{ i \in L_t^{(1)},\,\,j\in
      L_t^{(4)}, \, i\ne A,\,\,j\ne E,\,\,d_{iA}+d_{jE} < 2^t} (d_{iA}
  + 2^t + d_{jE})\nonumber\\
&+\sum_{i \in L_t^{(1)},\,\,j\in L_t^{(4)},\, i\ne C,\,\,j\ne
B,\,\,d_{iC}+d_{jB} < 2^t} (d_{iC} + 2^t + d_{jB})\nonumber\\
&+\sum_{i \in L_t^{(1)},\,\,j\in
      L_t^{(4)},\, i\ne A,\,\,j\ne E,\,\,d_{iA}+d_{jE} = 2^t}
  2^{t+1}\,.
\label{eq:app17}
\end{eqnarray}
The first terms equal the second ones and are denoted by $g_t$, and
the third term is denoted by $h_t$, so that $\Delta_t^{1,4} = 2g_t +
h_t$. One can compute the quantity $g_t$ as
\begin{eqnarray}
  g_t =& \sum_{m=1}^{2^t-2}\:\sum_{m^\prime = 1}^{2^t-1-m} \nu_t(m)\nu_t(m^\prime)(m+2^t+m^\prime)\nonumber\\
  =& \sum_{k=1}^{2^{t-1}-2}\:\sum_{k^\prime = 1}^{2^{t-1}-1-k} \nu_{t-1}(k)\nu_{t-1}(k^\prime)(2k+2^t+2k^\prime)\nonumber\\
  &+ \sum_{k=1}^{2^{t-1}-1}\:\sum_{k^\prime = 1}^{2^{t-1}-k} \nu_{t-1}(k)3^t (2k+2^t+2k^\prime-1)\nonumber\\
  &+ \sum_{k=1}^{2^{t-1}-1}\:\sum_{k^\prime = 1}^{2^{t-1}-k} 3^t \nu_{t-1}(k^\prime) (2k-1+2^t+2k^\prime)\nonumber\\
  &+ \sum_{k=1}^{2^{t-1}-1}\:\sum_{k^\prime = 1}^{2^{t-1}-k} 3^{2t}
  (2k-1+2^t+2k^\prime-1)\,.
\label{eq:app18}
\end{eqnarray}
The fourth terms can be summed directly, yielding
\begin{eqnarray}
3^{2t} 2^{t-3}(2^t-2)^2+3^{2t-1}2^{t-1}(2^{t-1}+1)(2^t-2).
\label{eq:app19}
\end{eqnarray}
In Eq.~(\ref{eq:app18}), the second and third terms are equal to
each other and can be simplified by first summing over $k^\prime$,
yielding
\begin{eqnarray}
&3^t\sum_{k=1}^{2^{t-1}-1} \nu_{t-1}(k) (3\cdot 2^{2t-2} - 2^t k -
k^2)\,. \label{eq:app20}
\end{eqnarray}
For use in Eq.~(\ref{eq:app20}), $\sum_{k=1}^{2^{t-1}-1}
\nu_{t-1}(k) = N_{t-1}-2$, and using Eq.~(\ref{eq:app15}),
\begin{eqnarray}
  \sum_{k=1}^{2^{t-1}-1} k \nu_{t-1}(k) &=
  \sum_{k=1}^{2^{t-1}} k \nu_{t-1}(k) - 2^{t-1}\nonumber\\
  &= 2^{t-3}(6^t-6)/5\,.
 \label{eq:app21}
\end{eqnarray}
Similarly to Eq.~(\ref{eq:app15}), we get
\begin{eqnarray}
&\sum_{k=1}^{2^{t-1}-1} k^2 \nu_{t-1}(k)=
\frac{1}{5}4^{t-1}6^{t-1}-\frac{1}{2}6^{t-1}+\frac{3}{10}4^{t-1}\,.
\label{eq:app22}
\end{eqnarray}
With these results, Eq.~(\ref{eq:app20}) becomes
\begin{eqnarray}
\frac{1}{2}3^t(2 \cdot4^{t-1}6^{t-1}-3 \cdot4^{t-1}+6^{t-1}) \,.
\label{eq:app23}
\end{eqnarray}
With Eqs.~(\ref{eq:app19}) and (\ref{eq:app23}),
Eq.~(\ref{eq:app18}) becomes
\begin{eqnarray}
   g_t =& 2g_{t-1} +
\frac{1}{6}3^t6^t4^t-\frac{1}{6}3^t6^t-\frac{3}{4}3^t4^t+2^{t-3}4^t9^t\nonumber\\
&-\frac{1}{2}4^t9^t+\frac{1}{2}2^t9^t\,. \label{eq:app24}
\end{eqnarray}
Considering the initial condition $g_1 = 0$, we can solve
Eq.~(\ref{eq:app24}) inductively leading to
\begin{eqnarray}
g_t =&
-\frac{3}{85}2^{t-3}(-171+17 \cdot2^{t+2}3^{t+1}+5 \cdot2^{t+3}3^{2t+1}\nonumber\\
&-85 \cdot9^t-17 \cdot4^{t+1}9^t)\,. \label{eq:app25}
\end{eqnarray}
To find an expression for $\Delta_t^{1,4}$, now the only thing left
is to evaluate $h_t$ as
\begin{eqnarray}
h_t &= 2^{t+1} \sum_{m=1}^{2^t-1} \nu_t(m) \nu_t(2^t-m)\nonumber\\
&= 2^{t+1} \sum_{m=1}^{2^t-1} \nu^2_t(m)\nonumber\\
&= 2^{t+1} \left[\sum_{k=1}^{2^{t-1}} 9^t  + \sum_{k=1}^{2^{t-1}-1}
  \nu^2_{t-1}(k)\right]\nonumber\\
&= 16^{2t} + 2h_{t-1}\,, \label{eq:app26}
\end{eqnarray}
where we have used the the symmetry $\nu_t(m) = \nu_t(2^t-m)$. Since
$h_1 = 36$, Eq.~(\ref{eq:app26}) is solved inductively:
\begin{equation}
h_t = \frac{9}{17}2^{t+1}(18^t-1)\,. \label{eq:app27}
\end{equation}
From Eqs.~(\ref{eq:app25}) and (\ref{eq:app27}),

\begin{eqnarray}
\Delta_t^{1,4} =&-&\frac{3}{85}2^{t-2}(-171+17 \cdot2^{t+2}3^{t+1}+5
\cdot2^{t+3}3^{2t+1}-85 \cdot9^t-17 \cdot4^{t+1}9^t)\nonumber \\
&+&\frac{9}{17}2^{t+1}(18^t-1)\,. \label{eq:app28}
\end{eqnarray}
Substituting Eqs.~(\ref{eq:app16}) and (\ref{eq:app28}) into
Eq.~(\ref{eq:app11}), we obtain the final expression for the
crossing paths $\Delta_t$:
\begin{eqnarray}
\Delta_t =
\frac{9}{25}2^{t-1}(2&+&3\cdot6^t)(14+6^{t+1})-\frac{9}{85}2^{t-1}(-171
+17 \cdot2^{t+2}3^{t+1}+5 \cdot2^{t+3}3^{2t+1}\nonumber\\&-&85
\cdot9^t -17 \cdot4^{t+1}9^t)+\frac{54}{17}2^{t+1}(18^t-1)-5
\cdot2^{t+1} \,, \label{eq:app29}
\end{eqnarray}
Substituting Eqs.~(\ref{eq:app29}) for $\Delta_m$ into
Eq.~(\ref{eq:app8}), and using $D_1 = 14$, we have
\begin{eqnarray}\label{eq:app9}
  D_t=\frac{1}{2200}(1243 \cdot2^t &-& 475 \cdot2^{t+2}3^t
+275 \cdot2^{t+3}3^t+275 \cdot2^t3^{2t+1}\nonumber\\&+&19
\cdot2^{3t+2}3^{2t+1} -11 \cdot3^{t+2}4^{t+1})
\end{eqnarray}
Inserting Eq.~(\ref{eq:app9}) into Eq.~(\ref{eq:app4}), one can
obtain the analytical expression for $\bar{d}_t$ in
Eq.~(\ref{eq:6}). 
\end{document}